\documentstyle[12pt,epsfig]{article}

\newcommand{\beq}{\begin{equation}}
\newcommand{\eeq}{\end{equation}}
\newcommand{\bea}{\begin{eqnarray}}
\newcommand{\eea}{\end{eqnarray}}

\newcommand{\gsim}{\lower.7ex\hbox{$
\;\stackrel{\textstyle>}{\sim}\;$}}
\newcommand{\lsim}{\lower.7ex\hbox{$
\;\stackrel{\textstyle<}{\sim}\;$}}

\def\lsim{\mathrel{\rlap{\lower3pt\hbox{\hskip0pt$\sim$}}
    \raise1pt\hbox{$<$}}}         %less than or approx. symbol
\def\gsim{\mathrel{\rlap{\lower4pt\hbox{\hskip1pt$\sim$}}
    \raise1pt\hbox{$>$}}}         %greater than or approx. symbol

\newcommand{\bibit}[1]{\bibitem{#1}}

\newcommand{\aver}[1]{\langle #1\rangle}

\newcommand{\La}{\overline{\Lambda}}
\newcommand{\Si}{\overline{\Sigma}}

\newcommand{\mhad}{\mu_{\rm hadr}}

\newcommand{\sigp}{\vec\sigma \vec\pi}

\newcommand{\GeV}{\,\mbox{GeV}}
\newcommand{\MeV}{\,\mbox{MeV}}
\newcommand{\matel}[3]{\langle #1|#2|#3\rangle}
\newcommand{\state}[1]{|#1\rangle}

\newcommand{\msp}[1]{\mbox{\hspace*{#1mm}~}}

\begin{document}
\thispagestyle{empty}
\vspace*{-20mm}

\begin{flushright}
Bicocca-FT-03-32\\
UND-HEP-03-BIG\hspace*{.08em}07\\
hep-ph/0312001
\vspace*{2mm}
\end{flushright}
\vspace*{15mm}

\boldmath
\begin{center}
{\LARGE{\bf
A `BPS expansion' for  $B$ and  $D$ mesons
}}
\vspace*{9mm} 

\end{center}

\unboldmath
\smallskip
\begin{center}
{\large{Nikolai~Uraltsev$^{*}$
}}  \\
\vspace{4mm}
{\sl Department of Physics, University of Notre Dame du Lac,
Notre Dame, IN 46556, USA}\\
{\small and} \\
{\sl INFN, Sezione di Milano, Milan, Italy} 
\vspace*{18mm}

{\bf Abstract}\vspace*{-.9mm}\\
\end{center}
\noindent
We analyze consequences of the approximation $\mu_\pi^2 \!-\!\mu_G^2
\!\ll\! \mu_\pi^2$ (a `BPS' limit) for $B$ and $D$ mesons. It is shown
that neglecting perturbative effects many power corrections would vanish to
all orders in $1/m_Q$, in particular those violating heavy {\sf
flavor} symmetry. Among them are corrections to $B\!\to\! D$
formfactors. A number of relations receive corrections only to the
second order expanding around the BPS limit to any order in $1/m_Q$, 
including both $f_+$ and $f_-$
at zero recoil. This allows an accurate evaluation of ${\cal F}_+\,$ for
$B\!\to\!D\,\ell \nu$. Its perturbative renormalization is computed
analytically in the required Wilsonian scheme, yielding the dominant
$3\%$ enhancement. 

\setcounter{page}{0}

\vfill

~\hspace*{-12.5mm}\hrulefill \hspace*{-1.2mm} \\
\footnotesize{ 
\hspace*{-5mm}$^*$On leave of absence from 
St.\,Petersburg Nuclear Physics 
Institute, Gatchina, St.\,Petersburg  188300, Russia}
\normalsize

\newpage

The heavy quark expansion is a powerful tool for treating strong
interactions relying on the first principles of QCD. Its
indispensable part is a general Wilsonian separation of effects
originating at sufficiently small and larger distances, a general
concept of the operator product expansion (OPE)
\cite{wilson,banda}. The resulting expansion is most
informative for inclusive decay probabilities where the OPE in local
heavy quark operators emerges \cite{vsope,buv}. %%% It turned out that 
Comprehensive
application of all elements of the heavy quark theory often allows to
put nontrivial constraints on the nonperturbative parameters in
individual transition amplitudes based on information from inclusive
decays. 

An important role here is played by the heavy quark sum rules,
including recently suggested spin sum rules \cite{sumrules} for the 
Small Velocity (SV) regime \cite{SV}. For example, they put a
rigorous bound $\varrho^2 \!\ge\! \frac{3}{4}$ on the slope of the IW function
and lead to a number of other inequalities in the heavy quark
limit. The recently proposed generalizations to higher orders in 
velocity \cite{orsay} resulted in remarkable relations (D'Orsay sum
rules) and nonperturbative bounds for its higher derivatives. 
The constraining power of the whole set of the SV sum rules   
depends strongly on
the actual size of the $B$ meson expectation value $\mu_\pi^2$ of the leading
local heavy quark nonperturbative kinetic operator $\bar{Q}(i\vec
D)^2Q$. Its theoretical expectations used to be a controversial subject
for years. The inequality between the expectation values of properly
defined kinetic and chromomagnetic operators puts a nontrivial lower
bound on $\mu_\pi^2$.

Although strong dynamics at small and large distances is governed by
the same QCD equations of motion, physics originating from below and
above a GeV scale is quite different. It has been pointed out \cite{chrom}
that experiment may implicate an interesting pattern for the
nonperturbative domain in $B$ and $D$ mesons, by favoring (see,
e.g.\ Ref.~\cite{DELPHI})
$\mu_\pi^2$ in the lower part of the allowed domain,\footnote{Certain
theoretical loopholes in the framework applied in individual analyses
have been identified \cite{amst,misuse,ckm03fpcp} which are expected to be
eliminated in the forthcoming experimental results.} only little
exceeding chromomagnetic expectation value
$\mu_G^2(1\GeV)\!=\!0.35^{+.03}_{-.02}\GeV^2$. In this case it is
advantageous to analyze nonperturbative dynamics combining the heavy
quark expansion with expanding around the point where
$\mu_\pi^2\!=\!\mu_G^2$ would hold. 

This is not just an arbitrary point of
a continuum in the parameter space, but a quite special limit where
the heavy flavor ground state has to satisfy functional relation
$\sigp\state{B}\!=\!0$, with $\sigma$ and $\pi$ being heavy quark spin
and momentum, respectively. From this perspective it is reminiscent to the
`BPS'-saturated state like the lowest Landau level for an electron in
magnetic field.\footnote{The analogy to this problem in quantum
mechanics and possible relation to BPS symmetry were first noted by
M.~Voloshin (1999, private communication), whom the suggested name of
the limit ascends to.} It is not clear how deep the analogy with BPS
symmetry goes,
for the Pauli Hamiltonian $\frac{(\sigp)^2}{2m_Q}$ describes
the leading-order power corrections rather than the static heavy quark
Hamiltonian itself
\beq
{\cal H}_\infty={\cal H}_{\rm light} - \int {\rm d}^3 {\bf x}\: 
Q^+ A_0 Q\,( {\bf x})
\label{10}
\eeq
shaping the ground state.
The functional relation would not be generally respected by
short-distance perturbative exchanges. It can only be viewed as an
approximate property of nonperturbative dynamics at energies below or
about $1\,$GeV scale, and applied to the ground state.

A practical consequence of the proximity to the BPS regime is small
room the heavy quark sum rules leave for possible values of the IW
slope 
\beq
\frac{3}{4} \lsim \varrho^2(1\GeV)\lsim 0.95\; \mbox{~~at~}
\mu_\pi^2(1\GeV) \lsim 0.45\GeV^2\;.
\label{12}
\eeq
A later UKQCD lattice evaluation \cite{rholat}\,
$\varrho^2\!=\!0.83^{+.15+.24}_{-.11-.01}$  fit well that prediction, although 
precision remains insufficient.
This and some other consequences were noted in Ref.~\cite{chrom}
(see also \cite{amst,ckm03fpcp}). In the present Letter we examine this
remarkable limit in more detail. We will largely abstract from the
perturbative corrections, in particular those renormalizing power
corrections to the heavy quark Lagrangian. To elucidate underlying
physics we often use quantum-mechanical language;
correspondence with second-quantized notations in field theory is
given, for instance in Ref.~\cite{optical}. The necessary introduction
into heavy quark expansion technique and used notations can be found
there, as well as in reviews \cite{rev,ioffe}. 
One of the practical
applications of the `BPS expansion' is an accurate model-independent 
determination of
$|V_{cb}|$ from the $B\!\to\! D\,\ell\nu$ rate near zero recoil if
enough statistics is available for this mode. Decays 
$\,B\!\to\! D\,\tau\nu_\tau$ sensitive to possible Higgs effects
add motivation for a precision control.
Power corrections to the
decay amplitudes in this kinematics are obtained through order $1/m_Q^2$.

\section{`BPS' relations for \boldmath $B$ and $D$ mesons}

The starting consequence of the BPS limit is equality of the
spin-singlet and spin-nonsinglet expectation values appearing in the
SV sum rules for asymptotically heavy quarks:
\beq
\varrho^2=\frac{3}{4}, \qquad \La=2\Si,\qquad \rho_{LS}^3=-\rho_D^3\;. 
\label{40}
\eeq
Similar relations hold for nonlocal correlators of the $1/m_Q$ terms
in the heavy quark Lagrangian:
\beq
\rho_{\pi G}^3=-2\rho_{\pi \pi}^3, \qquad \rho_{A}^3+\rho_{\pi G}^3=
-(\rho_{\pi \pi}^3+\rho_{S}^3), 
\label{42}
\eeq
a series extending to higher-order correlators. 

The BPS limit actually generalizes the heavy flavor symmetry to all
orders in $1/m_Q$. First, all terms $\propto 1/m_Q^k$ in the $1/m_Q$
expansion of the effective Hamiltonian annihilate the ground
state. Then all corresponding $T$-products likewise vanish in it, and
no power corrections to the heavy quark relation $M_P=m_Q+\La$
appear, so that 
\beq
M_B-M_D=m_b-m_c
\label{44}
\eeq
holds to all order in $1/m_Q$. It is also important that the 
Foldy-Wouthuysen transformation acts trivially (is unity) on the
ground state, therefore the proper nonrelativistic wavefunction
coincides with the upper component of the full Dirac bispinor in the
meson restframe. 

The above facts follow from the observation that for the `BPS'
wavefunction the lower component of the bispinor does not appear even when
power correction are included. To prove this, we recall the full QCD
equation of motion for the quark field in the nonrelativistic
notations. If the static solution $\varphi_0\,$ for $m_Q\!\to\!\infty$,
$\,\pi_0\varphi_0\!=\!0\,$ additionally satisfies the `BPS' constraint 
$\,(\vec\sigma \vec\pi)\varphi_0\!=\!0$, \,the bispinor 
\beq
Q(x,t)= e^{-im_Qt}\left(\begin{array}{c} \varphi_0(x,t) \\ 0 
\end{array} \right)
\label{46}
\eeq
solves the Dirac equation 
\beq
(\not\!\pi+m_Q\gamma_0)Q = m_Q \,Q\;, \qquad \pi_\mu\equiv iD_\mu-m_Q v_\mu\;, 
\label{48}
\eeq
and the corresponding wavefunction 
\beq
\Psi_\alpha(\vec{x}_Q,\{x_{\rm light}\}) \;=\;
\left(\begin{array}{cl} \Psi^0_\alpha (\vec{x}_Q,\{x_{\rm light}\})& 
\alpha\!=\!1,2\\ 0 & \alpha\!=\!3,4\end{array} \right)
\label{50}
\eeq
is a formal eigenstate with $E=m_Q+\La$ of the finite-$m_Q$
Hamiltonian including light degrees of freedom -- their wavefunction simply 
does not depend on $m_Q$. 

Absence of nontrivial Foldy-Wouthuysen corrections in the ground
state is also transparent. Since time derivative in the QCD Lagrangian
for fermions enters linearly with the gauge-field--independent
coefficient, the only source of the transformation to
nonrelativistic wavefunction is eliminating the lower component of the
bispinor. If the latter does not appear, Eq.~(\ref{46}), no
transformation is required. This corresponds to the fact that 
Foldy-Wouthuysen transformation $S_{{\rm FW}}$ (considered in the space
of the upper components, in the restframe) can be viewed as
\beq
S_{{\rm FW}}=\left(\frac{1+\gamma_0}{2}\right)^{-\frac{1}{2}}= 
1+\mbox{{\large $\frac{(\sigp)^2}{8m_Q^2}$}}+ 
\mbox{{\large $\frac{-\frac{1}{2}(\vec D\vec E_{_{}})+\frac{1}{2}\vec\sigma
\cdot\{\vec{E}\times \vec\pi-\vec\pi\times
\vec{E}\}}{8m_Q^3}$}} +...
\label{52}
\eeq
where the inverse square root of the projection operator is understood as the
local OPE expansion in $1/m_Q$ of the forward matrix element
$\bar{Q}\left(\frac{1+\gamma_0}{2}\right)^{\!-\frac{1}{2}\!}Q$ in terms of
the upper components. As 
shown in Ref.~\cite{opt2},
this expansion generates the whole nonrelativistic expansion for 
Dirac Hamiltonian, so triviality of Foldy-Wouthuysen
transformation is intimately related to absence of corrections to the
meson mass. 

Vanishing of corrections to heavy flavor symmetry in $1/m_Q$ expansion
applies also to the transition amplitudes between the ground-state
mesons with different heavy quarks, like in $B\!\to\!  D$ decays. Below we
take a closer look at the $B\!\to\! D$ amplitude assuming both $b$ and $c$
are heavy enough to meaningfully apply the expansion. Since axial 
current does not
contribute here, we focus on $\bar{c}\gamma_\mu b$-induced
amplitudes; the results apply to other allowed Lorentz
structures as well (say, scalar relevant for charge Higgs contributions). 

In general, the $B\!\to\! D$ amplitude is described by two vector
formfactors 
\beq
\matel{D(p_2)}{\bar{c}\gamma_\nu b}{B(p_1)}=
f_+(\vec{q}^{\,2})(p_1\!+\!p_2)_\nu + f_-(\vec{q}^{\,2})(p_1\!-\!p_2)_\nu\;.
\label{60}
\eeq
At zero recoil, $\vec{q}\!=\!0$ a single amplitude, viz.\ $\nu\!=\!0$ remains:
$J_0\!=\!(M_B\!+\!M_D)f_+(0)\!+\!(M_B\!-\!M_D)f_-(0)$. In the heavy 
quark limit one has
\beq
f_+(0)=\frac{M_B\!+\!M_D}{2\sqrt{M_B M_D}}, \qquad 
f_-(0)=-\frac{M_B\!-\!M_D}{2\sqrt{M_B M_D}}\;.
\label{62}
\eeq
It has been noted \cite{amst} that in the BPS limit all
power corrections to $J_0$ vanish. A stronger statement applies: the
$f_+$ formfactor determining all decays amplitudes with massless
leptons, at zero recoil keeps its asymptotic heavy quark limit value
\beq
f_+(0)=\frac{M_B+M_D}{2\sqrt{M_B M_D}}\;.
\label{64}
\eeq
Moreover, at arbitrary momentum transfer the heavy quark relation
between $f_+$ and $f_-$ remains valid in higher orders in $1/m_Q$:
\beq
f_-(q^2)=-\frac{M_B-M_D}{M_B+ M_D} f_+(q^2)\;.
\label{66}
\eeq
Furthermore, dynamic power corrections in $B\!\to\! D$ amplitude vanish:
\beq
f_+(q^2) = \frac{M_B\!+\!M_D}{2\sqrt{M_BM_D}} \;\,
\xi\!\left(\mbox{$\frac{M_B^2+M_D^2-q^2}
{2M_BM_D}$}
\right)\;,
\label{70}
\eeq
where $\xi(v_\mu^D v_\mu^B)$ is the normalized formfactor in the
infinite mass limit (IW function). Therefore, the $B\!\to\! D$
differential decay rate would more or less directly measure the IW
function. Below we show a way to derive these BPS consequences.

To relate $f_+$ and $f_-$ in Eq.~(\ref{66}) we can consider two
independent amplitudes, $\bar{c}\gamma_0 b$ and $i\partial_\mu J_\mu$;
applying the QCD equation of motion we have
\beq
i\partial_\mu J_\mu=(m_b\!-\!m_c)\:\bar{c}\,b\:, \qquad 
\matel{D}{i\partial_\mu J_\mu}{B}\!=\!(M_B^2\!-\!M_D^2)f_+ +
q^2 f_-\;.
\label{72}
\eeq
Both $\bar{c}\gamma_0 b$ and $\bar{c} b$ currents do not mix upper and
lower components for bispinors. Therefore, if the initial $B$ meson is
at rest and its lower component vanishes, the two currents coincide up
to the factor $m_b\!-m_c$ leading to
\beq
(m_b\!-\!m_c)\left[(M_B\!+\!E_D)f_+ +(M_B\!-\!E_D)f_- \right] =
(M_B^2\!-\!M_D^2)f_+ +(M_B^2\!+\!M_D^2\!-\!2M_B E_D)f_- \;.
\label{74}
\eeq
Replacing $\;m_b\!-\!m_c\;$ by $\;M_B\!-\!M_D\;$ we arrive at the
stated relation. Hence, it is a consequence of vanishing
lower components and of absent power corrections to the meson masses
in the BPS limit.

Non-renormalization of the (exact) zero-recoil amplitude can be
readily seen, for instance, in the heavy quark sum rule for $J_0$  extended to
arbitrary order in $1/m_Q$: the r.h.s.\ of the sum rule would not get
corrections, 
\beq
|F_D|^2+\sum_{\rm excit} |F_i|^2 =
1-\mbox{{\large
$\frac{\mu_\pi^2\!-\!\mu_G^2}{4}
\left(\frac{1}{m_c}\!-\!\frac{1}{m_b}\right)^2$}}
-\mbox{{\large $
\frac{\rho_D^3+\rho_{LS}^3}{4}\left(\frac{1}{m_c}\!+\!\frac{1}{m_b}\right)
\left(\frac{1}{m_c}\!-\!\frac{1}{m_b}\right)^2$}} - \;...\;,
\label{75}
\eeq
and inelastic transition amplitudes likewise vanish
order by order in the BPS limit. The simplest way to understand this,
however is using its quantum mechanical meaning revealed in
Ref.~\cite{opt6}.\footnote{The $1/m_Q^3$ term for the sum rule 
for $B\!\to\!D^*$
was derived in Ref.~\cite{rev}. The simplest way, in fact uses
this QM interpretation and explicit form of Foldy-Wouthuysen
transformation, Eq.~(\ref{52}).} 
We have seen that no corrections to the quantum
mechanical wavefunction appear, lower components are absent and there
is no explicit corrections in the $\bar{c}\gamma_0 b$ current -- then
no room for power corrections remains.

Absence of corrections to the heavy quark limit relation (\ref{70})
can be understood in the following way, looking once again at the
timelike component of the current. Assuming $B$ at rest and $D$ moving
with fixed velocity, we observe that the amplitude does not depend on
$m_b$ -- lower component of the $b$ field is absent, while the upper
one remains mass-independent. Therefore, it can only be a function of
$m_c$ and can be computed at $m_b\!\to\!\infty$. Changing the roles of $b$
and $c$ we would find it rather can only be a function of
$m_b$. Therefore it has to be free from any power corrections, at a given
mesons velocity.

Thus,
the heavy {\sf flavor} symmetry for the ground state would extend to any
order in $1/m_Q$ expansion in the strict BPS limit. 
\vspace*{3mm}

~~~~{\bf What if \boldmath $\;m_Q\!\to\! 0$\,?}\vspace*{1mm}

Absence of all power corrections to heavy quark relations even in a very
special regime may sound paradoxical. We definitely know that when
$m_Q\!\to\! 0$ usual $B$ and $D$ mesons would rather evolve to become
counterparts of the Goldstone particles like $\pi$ and $K$. The mass
relation for them would be clearly different, for instance. 

There is no contradiction between the two regimes even if we abstract
from the fact that the BPS limit cannot be exact and the corrections
to it analyzed in the $1/m_Q$ expansion would explode when $m_Q$
descends below a certain hadronic mass. The scale where the
BPS-protected heavy quark relations get completely destroyed may even
not decrease but remain stable when approaching the true BPS limit. For
power expansion cannot have a finite radius
of convergence being only asymptotic. Even if all power terms 
would vanish, there remain exponential terms scaling like \vspace*{-2mm}
\beq
\mbox{\Large$e^{-\frac{2m_Q}{\mhad}} $}
\label{82}
\eeq\vspace*{-8.5mm}

\noindent
which modify the heavy quark relations. 

Presence of such effects is a general feature of expansions in
field theory. In the heavy quark expansion they can explicitly appear
due to unlimited spectral density of the operator $\pi_0$ (heavy quark
Hamiltonian), with the support stretching below $-2m_Q$. Heavy
quark expansion excludes extra heavy quark degrees of freedom
expanding $1/(2m_Q\!+\!\pi_0)$ in $1/2m_Q$. The expansion is convergent
only if $|\pi_0|\!<\!2m_Q$ in the operator sense, which seems impossible
for a state well localized in space. 
Therefore, we should rather expect that violation of the BPS relations
becomes of order unity below a certain hadronic scale regardless of
proximity to the BPS regime for sufficiently heavy quarks. 

\thispagestyle{plain}
\begin{figure}[hhh]
\vspace*{-5mm} 
\begin{center}
 \mbox{\epsfig{file=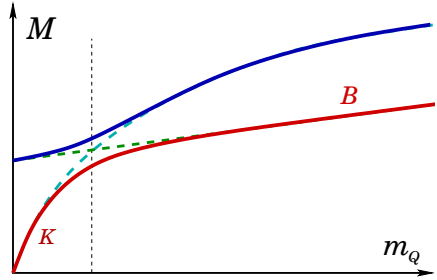,width=6cm}}
 \end{center}
\vspace*{-7mm}
\caption{ \small
~`Level crossing' would completely change $m_Q$-behavior below a
certain hadronic mass even at arbitrary weak ``BPS perturbation''}
\end{figure}

There are transparent mechanisms for such a change in the
regime. An obvious candidate is spontaneous breaking of the chiral
symmetry through nontrivial condensate $\aver{\bar{Q}Q}$ of the
$Q\bar{Q}$-pairs in the physical vacuum when the quark becomes light. The basic
assumption behind heavy quark symmetry, in contrast is separate conservation of
numbers of $Q$ and $\bar{Q}$. Even if a formal solution of equations
of motion exists, it may not be the correct ground state over the true
vacuum.

Another related mechanism is usual quantum mechanical level
crossing. If analytically continued eigenvalues for two states happen
to cross at some value of the heavy quark mass, the ``repulsion''
occurs, and behavior of the true ground state below and above the
crossing mass becomes totally different, while the effect of
any perturbation mixing the states becomes of order unity regardless of
its formal strength, see Fig.~1.

\section{Expanding around the BPS regime}

The BPS limit cannot be exact in actual QCD, rather an approximate
property of strong dynamics in the nonperturbative domain of momenta
below a GeV scale. Therefore, like with the conventional $1/m_Q$
expansion it is important to determine the scale of violation of its
particular predictions. If a concrete low-order in $1/m_Q$ correction
is found in terms of the heavy quark operators, it is not
difficult to see its BPS scaling. We, however, need a deeper
classification which would hold to all orders in $1/m_Q$.

The practical expansion parameter to quantify deviations from the
BPS limit, the norm of the state obtained by acting $\sigp$ on the
ground state 
\beq
\|\;\vec\sigma\vec\pi\,\state{B}\| = \sqrt{\mu_\pi^2\!-\!\mu_G^2}\;,
\label{90}
\eeq
has dimension of mass. A similar dimensionless parameter is
\beq
\beta=\|\pi_0^{-1}(\vec\sigma\vec\pi)\,\state{B}\| 
\equiv
\sqrt{3\!\left(\varrho^{2}\!-\!\mbox{$\frac{3}{4}$}\right)}=
3\!\left[\sum_n 
\;|\tau_{1/2}^{(n)}|^2\right]^{\!\frac{1}{2}}
\label{92}
\eeq
($\pi_0^{-1}$ in quantum mechanical notations is simply
$-\frac{1}{{\cal H}_{\infty} - \La}$ here). 
Numerically $\beta$ is not too small,
similar in size to generic $1/m_c$ expansion parameter in conventional
$1/m_c$ series. Relations violated to order $\beta$ may in practice be
more of a qualitative nature, while $\beta^2\!\propto\! 
\frac{\mu_\pi^2\!-\!\mu_G^2}{\mu_\pi^2}\:$ can provide enough
suppression. Moreover, we can count together powers of $1/m_c$ and
powers of $\beta$ to judge the real quality of a particular heavy
quark relation. 

BPS relations (\ref{66}), (\ref{70}) for the decay amplitude at
arbitrary recoil indeed receive corrections $\propto \!\beta^1$,
likewise equality of $\rho_{\pi G}^3$ and $-2\rho_{\pi\pi}^3$. Other
relations  between the heavy quark parameters mentioned in Sect.~1 are
accurate up to terms $\beta^2$. In particular, \\
$\bullet$ $M_B\!-\!M_D=m_b\!-\!m_c\:$ and $\:M_D=m_c+\La$.\\
$\bullet$ Zero recoil matrix element $\:\matel{D}{\bar{c}\gamma_0 b}{B}\:$
is unity up to $\beta^2$.\\
$\bullet$ Experimentally measured $B\!\to\!D\;$ formfactor $f_+$ near
zero recoil receives only second-order corrections in $\beta$ to all
orders in $1/m_Q$:
\beq
f_+\left( 0 \right) = \frac{M_B\!+\!M_D}{2\sqrt{M_B M_D}} 
+ {\cal O}(\beta^2)\;.
\label{94}
\eeq
This is an analogue of the Ademollo-Gatto theorem for the BPS
expansion. The similar statement clearly applies to $f_-$ as well. 

Absence of $\beta^{1}$ corrections to the meson mass is easily seen in
usual perturbation theory:
\beq
M_B=\matel{B}{{\cal H}_{\rm tot}}{B}\;.
\label{100}
\eeq
The terms linear in $\beta$ would reside either in $\state{B}$ or in
${\cal H}_{\rm tot}$. The former vanishes since to the leading,
$\beta^0$ order
$\state{B}$ remains the same BPS state at finite $m_b$. Absence of the
linear in $\beta$ terms from ${\cal H}_{\rm tot}$ is also understood
-- all local operators there include left-most $\sigp$ acting on
$\bar{b}$ and right-most $\sigp$ acting on $b$ since originate from
excluding the lower bispinor components; this is transparent in the
approach described in Ref.~\cite{opt2}.\footnote{In particular, this follows
from its Eq.~(10) once the scalar expectation value
$\matel{B}{\bar{b}b}{B}$ is $1+{\cal O}(\beta^2)$. The
energy-momentum tensor ${\cal D}$ does not depend explicitly on $m_b$
and plays the role of the rest-frame Hamiltonian of light degrees of freedom.}

Absence of linear non-BPS effects in zero-recoil
$\matel{D}{\bar{c}\gamma_0 b}{B}$ is also transparent in BPS
perturbation theory. Corrections $\,\propto \!\! \beta$ can originate either
from charm or from beauty wavefunction, but not from both
simultaneously,
\beq
\delta_{\beta^1}\matel{D}{\bar{c}\gamma_0 b}{B}= 
\langle\delta_{\beta^1}\Psi_D\state{\Psi_B^0} +
\langle\Psi_D^0\state{\delta_{\beta^1}\Psi_B} =0
\;,
\label{104}
\eeq
where we have used that the current acts on the unperturbed BPS state
as a unit operator except changing flavor. Alternatively, this follows
from the zero-recoil vector sum rule as described in the preceding
section -- power corrections to the sum rule appear only to order
$\beta^2$, and excitation transition {\sf amplitudes} appear to 
order $\beta$.

Vanishing of ${\cal O}(\beta)$ corrections in Eq.~(\ref{94}) for
$f_+(0)$ is least obvious. Since this holds for the combination
of $f_+$ and $f_-$ describing $J_0$, it suffices to show this for an
alternative combination, say taking the spacelike current $\bar{c}\vec\gamma
b$ in the case where one of the mesons is at rest and retaining only linear in
velocity terms. Once again we make use of the fact that 
order-$\beta^1$ effects could originate from perturbation in either $B$ or
$D$ wavefunctions, but not both simultaneously. To simplify algebra we
can assume that $B$ meson is at rest if BPS corrections appear in $B$,
and go to the $D$ restframe if a deviation from the BPS state occurs
in charm sector.

The next helpful observation is that if, say the beauty sector
wavefunction enters to the leading order in $\beta$, it can be assumed
to be in the more familiar $m_b\!\to\!\infty$ limit. Projecting the
matrix element on $\vec{v}$ we then get the spinor structure
\beq
\vec{n}\vec{J}\;=\; 2\sqrt{M_B M_D}\;
\left(\!\begin{array}{c} \varphi^{(c)} \\
\chi^{(c)}\end{array}\!\right)^{\!\dagger}
\left(\begin{array}{cc} 0 & \vec\sigma\vec{n}\\  \vec\sigma\vec{n} & 0
\end{array}\right)
\left(\begin{array}{c} \varphi^{(b)}_0(\vec{v}) \\
\!\frac{\vec\sigma\vec{v}}{2}
\,\varphi^{(b)}_0\end{array}\!\right),
\qquad \vec{n}=\frac{\vec{v}}{|\vec{v}\,|}\;,
\label{110}
\eeq
where $\varphi^{(c)}$ and $\chi^{(c)}$ refer to $D$ at rest and
include all mass corrections. 

The term $\matel{\varphi^{(c)}}{(\vec\sigma\vec{n}) \frac{\vec\sigma
\vec{v}}{2}}{\varphi^{(b)}_0}$ yields just the heavy quark limit
result, Eq.~(\ref{62}) if $\varphi^{(c)}$ coincided with the
asymptotic wavefunction $\varphi^{(c)}_0$. It generally does not, but 
the first-order corrections
to the overlaps like $\langle\varphi_0^{(c)}\state{\varphi_0^{(b)}}$ 
always vanish.
This applies to BPS perturbations since $\varphi^{(c)}$
remains normalized to unity up to terms $\propto \!\beta^2$ to any order
in $1/m_c$. 

Another source of power corrections is the product 
$\matel{\chi^{(c)}}{(\vec\sigma\vec{n})}{\varphi^{(b)}_0(\vec{v})}$. 
Now we use that $\varphi^{(b)}_0(\vec{v})$ is the asymptotic heavy
quark wavefunction which velocity dependence to the first order is
given by 
\beq
\state{\varphi^{(b)}_0(\vec{v})}=
\state{\varphi^{(b)}_0(0)}-\frac{1}{{\cal H}_\infty \!-\!\La}\; 
\vec\pi \vec{v} \,\state{\varphi^{(b)}_0(0)} \equiv
\state{\varphi^{(b)}_0(0)}+ \pi_0^{-1\,}\vec\pi \vec{v}\, 
\state{\varphi^{(b)}_0(0)} \,,
\label{112}
\eeq
where ${\cal H}_\infty$ is the static heavy quark limit Hamiltonian,
and the second form uses the analogue of the second-quantized
notations. Since heavy quark spin $\vec\sigma$ commutes with ${\cal
H}_\infty$ this piece reads
\beq
\mbox{$\frac{|\vec{v}|}{3}$}\: 
\matel{D}{\,\bar{c}\,\mbox{$\frac{1\!-\!\gamma_0}{2}$} \pi_0^{-1}
(\vec\sigma\vec\pi)\, b\,}{B_0(\vec{v})}
\label{114}
\eeq
when calculated for spinless meson states. Therefore it vanishes
for the BPS beauty state regardless of the charm wavefunction.

The above reasoning essentially relies on linear in $\vec{v}$
approximation. To order $\vec{v}^{\,2}$, for instance, the operator
$\vec{\pi}^{\,2}$ appears along with $(\vec\sigma\vec\pi)^2$, which does
not vanish when acts on the BPS state. Therefore, while having managed to
leave the point of zero recoil, we cannot extend the
Ademollo-Gatto theorem to, say the slope of the formfactor. 

\section{Application to \boldmath $\,B\!\to\!D\,\ell\nu\,$ near zero recoil}

Since $f_-$ formfactor does not contribute to any decay amplitude with
massless leptons, the amplitude even near zero recoil depends on the
space-like current which suffers from linear $1/m_c$ effects in the
$1/m_Q$ expansion; conventionally this is viewed as the
serious theoretical drawback of such a decay mode. The BPS expansion
turns out more robust in this respect protecting the heavy quark
relation up to the second order. This allows an accurate estimate of the
$B\!\to\!D\,$ rate near zero recoil in terms of $\,|V_{cb}|^2$. 

Similar to heavy quark symmetry itself, the BPS limit is affected by
short-distance perturbative effects. Accounting for the latter should
therefore be done accordingly. The Wilsonian procedure with the
explicit `hard' cutoff $\mu$ is most suitable here simply eliminating the
low-momentum domain.
The principal perturbative corrections are just short-distance 
renormalization of the bare $\bar{c}\gamma_\nu b$ current itself. The
technique for such Wilsonian calculations has been elaborated and
applied to zero-recoil $B\!\to\! D^*$ amplitude, see
Ref.~\cite{ioffe}. The method specifically corresponds to the adopted
renormalization scheme for nonperturbative operators with the upper
cutoff $\mu$ on energy \cite{chrom}. The case of
$B\!\to\!D$ assuming non-vanishing velocity requires only technical
modifications.

The one-loop result with arbitrary cutoff scale $\mu$ is obtained in
an analytic form: 
{\small
$$
\mbox{{\normalsize $\xi_V(\mu) \!=\! 1 + $}}\frac{2\alpha_s}{3\pi}
\left[
\frac{3m_b^2\!+\!2m_c m_b\!+\!3m_c^2}{2(m_b^2\!-\!m_c^2)} 
\ln{\frac{\mu\!+\!\omega_b}{\mu\!+\!\omega_c}} - 2
- \mu \left\{\frac{4}{3\mu^2} 
\frac{m_c\omega_b\!-\!m_b\omega_c}{m_b\!-\!m_c}
+\frac{2}{3}\frac{\frac{m_c}{\omega_b}\!-\!\frac{m_b}{\omega_c}}{m_b\!-\!m_c}
\right.\right. \msp{13}
$$
$$
\msp{25}
\left.\left.
- \frac{1}{3}\, 
\frac{\frac{\omega_b}{m_b} \!-\!\frac{\omega_c}{m_c}}{m_b\!-\!m_c}
\!+\!\frac{2m_c m_b}{m_c\!+\!m_b} \,
\frac{\frac{1}{\omega_b}\!-\!\frac{1}{\omega_c}}{m_b\!-\!m_c}
\!+\! \frac{1}{6}\,\frac{1}{m_c\!+\!m_b}\left(
\frac{\omega_c}{m_c}\left(3\!-\!\frac{m_b}{m_c}\right)+
\frac{\omega_b}{m_b}\left(3\!-\!\frac{m_c}{m_b}\right)
\right)
\right.\right.
$$
\beq
\msp{5}
\left.\left.
+ \frac{2}{3} \frac{2m_c m_b}{m_c\!+\!m_b} 
\left(\frac{m_b}{\omega_b^3} \!+\!\frac{m_c}{\omega_c^3}\right)
\!+\!\frac{\mu}{6}\left(\frac{1}{m_c} \!-\!\frac{1}{m_b}\right)^2
\!-\!\frac{2}{3}\frac{\mu^2}{m_c m_b} \frac{\frac{m_b^2}{\omega_c}
\!+\!\frac{m_c^2}{\omega_b}}{m_b^2\!-\!m_c^2}
\right\}
\right]\;,
\label{150}
\eeq
}
where
\beq
\omega_c= \sqrt{m_c^2+\mu^2}, \qquad \omega_b= \sqrt{m_b^2+\mu^2}\;.
\label{152}
\eeq
Fig.~2 shows it numerically assuming $m_c\!=\!1.2\GeV$,  $m_b\!=\!4.6\GeV$ and
$\alpha_s\!=\!0.3$.

\thispagestyle{plain}
\vspace*{-3mm}
\begin{figure}[hhh]
 \begin{center}
 \mbox{\epsfig{file=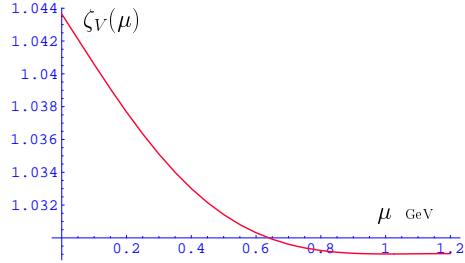,width=6cm}}
 \end{center}
\vspace*{-6mm}
\caption{ \small
Short-distance renormalization for $B\!\to\!D$ amplitude at
$m_c\!=\!1.2\GeV$,  $m_b\!=\!4.6\GeV$ and $\alpha_s\!=\!0.3$ as a 
function of the separation scale $\mu$.}
\end{figure}

The first-order $1/m_Q$ correction is readily read off the current in
Eq.~(\ref{110}), cf.\ Eq.~(\ref{114}):\footnote{A connection between
the $1/m_Q$ correction and inclusive transitions was first noted in
Ref.~\cite{orsayisr}.}
\beq
{\cal F}_+ =\frac{2\sqrt{M_B M_D}}{M_B+M_D}\; f_+(0)=
\xi_V(\mu) + 
\left(\frac{\La}{2}\!-\!\overline\Sigma\right) 
\left(\frac{1}{m_c}\!-\!\frac{1}{m_b}\right)
\frac{{M_B\!-\!M_D}}{M_B\!+\!M_D}-
{\cal O}\left(\!
\frac{1}{m_Q^2}\!\right)\;;
\label{160}
\eeq
it is positive, explicitly suppressed by two powers of 
$\beta$ in accord with
the theorem in the previous section, and is small numerically. One
finds that linear $\mu/m_Q$ dependence of one-loop $\xi_V(\mu)$
cancels against the one-loop $\mu$-dependence 
of $\frac{\La}{2}\!-\!\overline\Sigma$, as expected. 

Although $\frac{\La}{2}\!-\!\overline\Sigma$ normalized at $0.8\mbox{
to }1\GeV$ has not been directly measured, a good estimate for it
would be the actual upper bound \cite{ioffe}
\beq
\frac{\La}{2}\!-\!\overline\Sigma =\frac{1}{3}\,\frac{1}{2M_B}
\matel{B}{\bar{b}(\sigp)(-\pi_0^{-1})(\sigp) b }{B}
 \;\le\;
\sqrt{\left(\varrho^2\!-\!\mbox{$\frac{3}{4}$}\right)\, 
\frac{\mu_\pi^2\!-\!\mu_G^2}{3}}
\;.
\label{162}
\eeq
The IW slope $\varrho^2$ has not been well determined
experimentally yet. An alternative estimate would use the scale of
an average excitation energy $\tilde\varepsilon$ of the $P$-wave
$\frac{1}{2}$-states saturating the sum rules:
\beq
\frac{\La}{2}\!-\!\overline\Sigma \;\approx\;
\frac{\mu_\pi^2-\mu_G^2}{3\tilde\varepsilon}
\label{164}
\eeq
with  $\tilde\varepsilon\approx 500 \;\mbox{to}\;700\MeV$. Depending
on the precise value of $\mu_\pi^2(1\GeV)$ the $1/m$ correction to
${\cal F}_+$ emerges at $1\%$ level. 

Since the expansion in $1/m_c$ is involved, higher-order corrections {\it a
priori}\, could be significant. This is illustrated by the perturbative
contribution where linear in $\mu$ approximation works well only for
$\mu \!\lsim\! 400\MeV$, while already at $\mu \!\simeq\! 800\MeV$ second- and
third-order terms are of the same size. The BPS expansion
ensures, however that the overall suppression is carried on to all
higher-order nonperturbative effects as well. 

The second-order power correction to ${\cal F}_+$ has three
pieces, each manifestly of the second order in BPS. It also follows
directly from Eq.~(\ref{112}), now explicitly expanding in
$1/m_c$:
\beq
\delta_{1/m^2} {\cal F}_+ = \left(\frac{1}{m_c}\!-\!\frac{1}{m_b}\right)^2
\left\{V_1+V_2+V_3\right\}
\label{167}
\eeq
The first one, $V_1$ is the same 
as in zero-recoil
$\matel{D}{\bar{c}\gamma_0}{B}$ and is well understood
\cite{optical}. It is negative consisting of local correction and
nonlocal ``overlap deficit'': 
$$
V_1= -\frac{\mu_\pi^2\!-\!\mu_G^2}{8}
\cdot \left(1+\chi^{(V)}_{\mbox{\footnotesize n-l}}\right), \qquad\qquad 
(\mu_\pi^2\!-\!\mu_G^2) \,\chi^{(V)}_{\mbox{\footnotesize n-l}}=
\frac{1}{2M_B}\matel{B}{\bar{b}(\sigp)^2\pi_0^{-2}(\sigp)^2 b }{B} 
$$
\beq
\msp{30}
= \; \int \!i|x_0|\,{\rm d}^4x\, \frac{1}{4M_{B}}
\matel{B}{\,iT\{\bar{b}(\sigp)^2 b(x),\, \bar{b}(\sigp)^2
b(0)\}\,}{B}'
 >0\;.
\label{168}
\eeq
The second piece is local and positive, 
$V_2\!=\!\frac{\mu_\pi^2\!-\!\mu_G^2}{6}$. 
The last piece can be viewed as the $1/m_Q$ correction to the value
of $\frac{\La}{2}\!-\!\overline\Sigma$ in a finite-mass 
meson and is 
described by the expectation value of the nonlocal correlator 
\beq
-6V_3= \frac{1}{2M_B}\matel{B}{\bar{b}(\sigp)^2\pi_0^{-1}(\sigp)
\pi_0^{-1}(\sigp)b }{B}^\prime\;.
\label{169}
\eeq
It is plausible that the correlator is
likewise positive decreasing the effective value of
$\frac{\La}{2}\!-\!\overline\Sigma$, although strictly speaking the
sign is not fixed. Its scale can be roughly estimated as that for the
correlator $\bar{Q}(\sigp) \pi_0^{-1}(\sigp)^2 \pi_0^{-1}(\sigp) Q$ and
taking for the latter $\frac{\mu_\pi^2}{\tilde\varepsilon}\,
\left(\frac{\La}{2}\!-\!\overline\Sigma \right)$ as an educated
dimensional guess. Since local pieces nearly cancel, $\delta_{1/m^2}
{\cal F}_+$ is dominated by two nonlocal correlators.

Collecting all terms through the second order in $1/m_Q$ we arrive at 
the estimate
\beq
\delta_{\rm power\,} {\cal F}_+ \lsim 0.01 \; 
\mbox{~at~} \mu_\pi^2 \lsim 0.43\GeV^2 
\;,
\label{176}
\eeq
%%%through the second order in $1/m_Q$, 
where following
Refs.~\cite{optical,vcb} we assume $\chi^{(V)}_{\mbox{\footnotesize
n-l}}\!=\!0.5\pm 0.5$. The literal prediction depends moderately on the actual
kinetic expectation value,
\beq
{\cal F}_+ \simeq 1.04 + 0.13\, 
\frac{\mu_\pi^2(1\GeV)\!-\!0.43\GeV^2}{1\GeV^2}\;;
\label{178}
\eeq
the uncertainty, however would soon go out of control at this level of
precision for $\mu_\pi^2(1\GeV)$ exceeding $0.45\GeV^2$.

The above estimates suggest that the 
nonperturbative corrections to ${\cal F}_+$ are really tiny, 
and it can be accurately
evaluated. It should be appreciated, however that practical
implementations of the heavy quark expansion leave out possible
exponential terms like
\beq
\delta_{\mbox{\footnotesize exp}\,}{\cal F}_+ \propto 
\left(e^{-\frac{m_c}{\mhad}} - e^{-\frac{m_b}{\mhad}}\right)^2
\label{182}
\eeq
which are routinely ignored, but may be essential at a percent level
when relying on heavy charm expansion \cite{vadem}. At the moment we
cannot say much about them; yet they probably put the actual limit on
the accuracy of the theoretical predictions for ${\cal F}_+$ if
$\mu_\pi^2(1\GeV)$ is confirmed to be below $0.45\GeV^2$. 

\section{Conclusions}

We have analyzed the structure of the pure nonperturbative corrections
to $B$ and $D$ mesons near the `BPS' regime which would apply if 
nonperturbative physics is largely confined below $1\GeV$ scale and
yield $\mu_\pi^2$ close to $\mu_G^2$. A number of relations are shown
to hold extending the heavy flavor (but not spin!) symmetry to all
orders in $1/m_Q$ in the BPS limit, valid however only for the ground
state pseudoscalar heavy flavor mesons. Some of the important BPS
relations get corrections only to the second order in the BPS
expansion  regardless of the order in $1/m_Q$, among them are two
$B\!\to\!D$ zero recoil formfactors. This is the analogue of the
Ademollo-Gatto theorem for the BPS expansion. 

The practical utility of such an expansion strongly depends on the
actual size of $\mu_\pi^2(1\GeV)$. If -- as has been reported by most
experimental analyses -- it centers at or below $0.4$, or even up to
$0.45\GeV^2$, it is rather powerful. At larger values of $\mu_\pi^2$
the predictability weakens; in this case, however any application
relying on the heavy quark expansion for charm for non-BPS--protected
relations would become questionable. 

Dynamic heavy quark expansion has enlightened us that the scale of
nonperturbative effects for heavy quarks in actual QCD is quite
significant, $\mu^{\rm np}\!\gsim\! \sqrt{\mu_\pi^2}\!\simeq\! 700\MeV$ at
least. This contrasted early ideas that this scale is like a
`constituent' light quark mass $m_{\rm const}\!\sim\! 250\MeV$ inherited
from naive nonrelativistic quark models. Yet it turns out that the
latter small scale can sometimes re-emerge in theory -- where the
`BPS'-protected properties are considered, with 
$\sqrt{\mu_\pi^2\!-\!\mu_G^2} \!\lsim \!300\MeV$. 
A better theoretical understanding of the
dynamic origin of such a hierarchy is highly desirable.

An interesting place to test the BPS predictions is the ratio 
$\,f_-(0)/f_+(0)\,$ fixed in the heavy quark limit,
\beq
\frac{f_-(0)}{f_+(0)}=-\frac{M_B\!-\!M_D}{M_B\!+\!M_D}\;.
\label{210}
\eeq
Since both $f_+(0)$ and $f_-(0)$ undergo power corrections only to the
second order in the BPS expansion, they must be small and accurately
evaluable, Sect.~3, thereby offering a probe for possible exponential
effects. The suppressed formfactor $f_-$ can be measured in the
$B\!\to\!D\,\tau\nu_\tau $ decays in future high-statistics
experiments. This mode also probes possible effect of charge Higgs
exchanges, which requires a precision understanding of the SM
amplitude. The BPS expansion provides support for such a treatment;
the electroweak corrections are to be properly incorporated at this level.

The heavy quark expansion together with heavy quark sum rules allows
an accurate prediction for the $B\!\to\!D$ amplitudes near zero recoil
in terms of measured observables. Our estimate (the electroweak 
effects, in particular the factor of $1.007$ from the 
universal short-distance renormalization 
are not included here) depends on
$\mu_\pi^2(1\GeV)$, but for moderate values the power corrections appear
at a percent level:
\beq
{\cal F}_+ = 1.04 \pm 0.01_{\rm power} \pm 0.01_{\rm pert} +
\delta_{\mbox{\footnotesize exp}}\;,
\label{216}
\eeq
with perturbative corrections from momenta above $1\GeV$ contributing
the dominant piece of $3\%$. It can be further refined. The 
corrections are significantly smaller and more definite compared to
the `gold plated' $B\!\to\!D^*$ decay mode.

This assessment differs from the existing estimates \cite{earlyest}
although, in principle is compatible with the predictions within 
their respective large
error bars. The rationale is readily seen equating and counting 
together orders in conventional $1/m_c$ and in the `BPS' expansion:
Eq.~(\ref{160}) gives power corrections through the {\bf third} order,
with estimates included for the fourth-order effects. In contrast, no
BPS-backup exists for $B\!\to\!D^*$, the corrections are significant 
in the BPS regime \cite{chrom}, and they are uncertain already 
to the leading order $1/m_c^2$.

\vspace*{2mm}

{\bf Acknowledgments:} I am grateful to I.~Bigi, M.~Shifman and in
particular to A.~Vainshtein for interest, illuminating discussions and
for sharing a nontrivial alternative perspective.
This work was supported in part by the NSF under grant number PHY-0087419.

\end{document}